\documentclass[aps,pra,twocolumn,10pt,superscriptaddress,nofootinbib,balancelastpage]{revtex4-1} 
\usepackage[latin1]{inputenc}
\usepackage{amsmath,amssymb}
\usepackage{mathrsfs} 
\usepackage[capitalise]{cleveref}
\usepackage{braket}
\usepackage{graphicx,color,colortbl}
\usepackage{booktabs}
\usepackage{dsfont}

\allowdisplaybreaks

\newcommand{\dif}{\mathrm{d}}%
\newcommand{\Nabla}{\vec{\nabla}}%
\newcommand{\Laplace}{\boldsymbol{\triangle}}%
\newcommand{\norm}[1]{\lVert#1\rVert}%
\newcommand{\Kronecker}[2]{\delta_{#1#2}}%
\newcommand{\ZT}[1]{\textquotedblleft#1\textquotedblright}%

\newlength{\myl}%
\newcommand{\INT}[3]{\settowidth{\myl}{$\displaystyle\int_{#1}^{#2}$}{\int_{#1}^{#2}\;\;\;\hspace{-\the\myl}\dif #3}\,}
\newcommand{\TINT}[3]{\settowidth{\myl}{$\displaystyle\int_{#1}^{#2}$}{\int_{#1}^{#2}\;\;\;\;\;\,\hspace{-\the\myl}\dif #3}\,}
\newcommand{\EINT}[3]{\settowidth{\myl}{$\int_{#1}^{#2}$}{\int_{#1}^{#2}\;\;\;\,\hspace{-\the\myl}\dif #3}\,}

\makeatletter
\newcommand\footnoteref[1]{\protected@xdef\@thefnmark{\ref{#1}}\@footnotemark}
\makeatother

\begin{document}

\title{Predictive local field theory for interacting active Brownian spheres in two spatial dimensions}

\author{Jens Bickmann}
\affiliation{Institut f\"ur Theoretische Physik, Westf\"alische Wilhelms-Universit\"at M\"unster, D-48149 M\"unster, Germany}
\affiliation{Center for Soft Nanoscience, Westf\"alische Wilhelms-Universit\"at M\"unster, D-48149 M\"unster, Germany}

\author{Raphael Wittkowski}
\email[Corresponding author: ]{raphael.wittkowski@uni-muenster.de}
\affiliation{Institut f\"ur Theoretische Physik, Westf\"alische Wilhelms-Universit\"at M\"unster, D-48149 M\"unster, Germany}
\affiliation{Center for Soft Nanoscience, Westf\"alische Wilhelms-Universit\"at M\"unster, D-48149 M\"unster, Germany}
\affiliation{Center for Nonlinear Science, Westf\"alische Wilhelms-Universit\"at M\"unster, D-48149 M\"unster, Germany}

\begin{abstract}
We present a predictive local field theory for the nonequilibrium dynamics of interacting active Brownian particles with a spherical shape in two spatial dimensions. The theory is derived by a rigorous coarse-graining starting from the Langevin equations that describe the trajectories of the individual particles. For maximal accuracy and generality of the theory, it includes configurational order parameters and derivatives up to infinite order. In addition, we discuss possible approximations of the theory and present reduced models that are easier to apply. We show that our theory contains popular models such as Active Model B + as special cases and that it provides explicit expressions for the coefficients occurring in these and other, often phenomenological, models. As a further outcome, the theory yields an analytical expression for the density-dependent mean swimming speed of the particles. To demonstrate an application of the new theory, we analyze a simple reduced model of the lowest nontrivial order in derivatives, which is able to predict the onset of motility-induced phase separation of the particles. By a linear stability analysis, an analytical expression for the spinodal corresponding to motility-induced phase separation is obtained. This expression is evaluated for the case of particles interacting repulsively by a Weeks-Chandler-Anderson potential. The analytical predictions for the spinodal associated with these particles are found to be in very good agreement with the results of Brownian dynamics simulations that are based on the same Langevin equations as our theory. Furthermore, the critical point predicted by our analytical results agrees excellently with recent computational results from the literature.
\end{abstract}
\maketitle

\section{Introduction}
Active Brownian particles (ABPs), which combine Brownian motion and propulsion, are an important type of active matter that is currently attracting great scientific interest \cite{Romanczuk2012,Wensink2013,MIPS,Elgeti2015,BechingerEA16,Fodor16,Speck2016,Zttl2016,Marconi2017,Mallory18}. 
Both artificial self-propelled microparticles \cite{Rao2015,Wu2016,Xu2016,Guix2018,Chang2019,PachecoJerez2019} and motile microorganisms \cite{SchwarzLinek2016,Chen2017,Andac2019} are frequently described as ABPs. Even bacteria like \textit{Escherichia coli}, which show a run-and-tumble motion \cite{Berg2008,Tailleur2008,Paoluzzi_2013,Liang2018}, are often successfully modeled as ABPs \cite{Tailleur2008,Cates2013,Liu2017,Andac2019}.
Due to their self-propulsion, already the common simple ABPs with a spherical shape exhibit a variety of unusual effects like accumulation at nonattracting walls \cite{Elgeti2015,BechingerEA16,Duzgun2018,Das2019}, superfluidity \cite{Takatori17}, anomalous Casimir forces \cite{Ni15}, negative interfacial tension \cite{NegativeInterfaceTensionBialk15}, reversed Ostwald ripening \cite{Tjhung2018}, non-state-function pressure \cite{Solon2015Nat,Solon2015a}, and motility-induced phase separation (MIPS) \cite{MIPS}. The latter effect originates from the complex nonequilibrium dynamics of interacting ABPs and gained particularly strong scientific attention in recent years \cite{Tailleur2008, Fily2012, Bialk2013, Buttinoni2013, Redner2013, Stenhammar2013, Speck2014, Wittkowski2014, Wysocki2014, Zoettl2014, Solon2015a, Redner2016, RW, Digregorio2018, Paliwal2018, Solon2018, Whitelam2018,Nie2019}. 

A powerful tool for investigating the collective behavior of ABPs are field theories. While particle-based computer simulations were the dominant approach in the past research on ABPs, field-theoretical approaches are relatively rare, although they often allow deeper insights into the properties of an active system via the underlying equations. The existing field theories for ABPs include nonlocal as well as local ones. Nonlocal field equations can be more compact and can capture certain properties of the described system more appropriately, but they are typically much more difficult to interpret and to treat numerically than corresponding local field equations. Therefore, most of the available field theories for ABPs are local.   
An example for existing nonlocal field theories for ABPs are generally active dynamical density functional theories \cite{Rex2007, Wittkowski2011,Menzel2016j}. These theories, however, are limited to weak propulsion and, when they involve too strong approximations, also to low particle concentrations.   
The existing local field theories for ABPs include phase field crystal (PFC) models \cite{EmmerichLWGTTG12,Menzel2013,Menzel2014,Alaimo2016,Alaimo2018, Praetorius2018}. These models can be derived from dynamical density functional theories and their applicability is  therefore similarly limited to close-to-equilibrium systems. 
In addition, there is a number of individual models for ABPs including active diffusion equations \cite{Cates2013,Bialk2013}, an extension towards mixtures for active and passive Brownian particles \cite{RW}, a model with an explicit particle-field representation based on the concept of particle-wave duality \cite{Grossmann19}, a hydrodynamic model including the flow field of ABP suspensions \cite{Steffenoni17}, Cahn-Hilliard-like models \cite{Stenhammar2013,Speck2014}, the related nonintegrable Active Model B (AMB) \cite{Wittkowski2014}, and its extension Active Model B + (AMB+) \cite{Tjhung2018,cates_tjhung_2018}.
To keep the models relatively simple, they involve strong approximations and often only terms of the lowest nontrivial order in the order-parameter fields and derivatives. This, however, reduces their applicability and accuracy.

In this article, we present a highly general and accurate local field theory for the nonequilibrium dynamics of interacting ABPs. As in the most existing simulation studies on ABPs, we focus on spherical particles without hydrodynamic interactions in two spatial dimensions. To obtain a predictive theory, where all parameters of the field equations are given by explicit expressions that relate them to the microscopic properties of the considered system, the theory is derived via a rigorous coarse-graining starting at the commonly used Langevin equations describing the motion of individual ABPs. 
For high applicability and accuracy, approximations are kept to a minimum. In its initial form, the theory therefore takes order parameters and derivatives up to infinite order into account. On this basis, we present systematic approximations that lead to reduced models with finite field equations of the wanted complexity.  
Comparing our theory with the aforementioned local models from the literature, we show that all models that consider the same type of systems can be identified as special cases of our theory.
We also use our theory to derive an analytical expression for the density-dependent mean swimming speed in a homogeneous system of spherical ABPs. 
Furthermore, the theory provides an analytical expression for the spinodal corresponding to the onset of MIPS in a system of ABPs, where the interaction potential can be specified by the user. The theoretical predictions for the spinodal and especially the critical point are found to be in excellent agreement with recent simulation results \cite{Siebert2018,Jeggle2019}.
 
The article is structured as follows: In section \ref{sec:derivation}, the general field theory is derived and possible approximations are provided. 
On this basis, in section \ref{sec:MODELS} a set of reduced models is derived and compared to other existing models from the literature. Examples for applications of the theory are demonstrated in section \ref{sec:Applications}. Finally, concluding remarks are given in section \ref{sec:Conclusion}.

\section{\label{sec:derivation}Derivation of the general field theory and approximations}
\subsection{\label{ssec:gfe}General field theory}
The ABP system most commonly considered in previous studies is given by $N$ similar active Brownian spheres that can translate in a horizontal plane and rotate about vertical axes through the particles' centers. Their motion originates from an underlying Brownian motion, the persistent self-propulsion of the particles, and interactions between them.  
The motion of such particles can be described by their center-of-mass positions $\{\vec{r}_i(t)\}$ and orientations $\{\phi_i(t)\}$ as functions of time $t$, where the index $i\in\{1,\dotsc,N\}$ distinguishes the individual particles. 
Suitable equations of motion for the ABPs are given by the overdamped Langevin equations
\cite{Fily2012,Bialk2013,Buttinoni2013,Redner2013,Speck2014,Ni15,Solon2015a,Redner2016,Speck2016,NegativeInterfaceTensionBialk15,RW,Digregorio2018,Duzgun2018,Siebert2018,Tjhung2018, Jeggle2019}
\begin{align}
\dot{\vec{r}}_i &= \vec{\xi}_{\mathrm{T},i} + v_0\hat{u}(\phi_i) + \beta D_\mathrm{T} \vec{F}_{\mathrm{int}, i}(\{\vec{r}_i\}), \label{eqn:LangevinR}\\
\dot\phi_i &= \xi_{\mathrm{R},i}. \label{eqn:LangevinPHI}%
\end{align}
Here, a dot over a variable denotes a partial derivative with respect to time. The translational and rotational Brownian motion of the $i$-th particle is described by statistically independent Gaussian white noises  $\vec{\xi}_{\mathrm{T}, i}(t)$ and $\xi_{\mathrm{R}, i}(t)$, respectively. Their correlations are given by $\braket{\vec{\xi}_{\mathrm{T}, i}(t_1)\otimes\vec{\xi}_{\mathrm{T}, j}(t_2)} = 2D_\mathrm{T}\Kronecker{i}{j}\mathds{1}_2\delta(t_1-t_2)$ and $\braket{\xi_{\mathrm{R}, i}(t_1)\xi_{\mathrm{R}, j}(t_2)} = 2D_\mathrm{R}\Kronecker{i}{j}\delta(t_1-t_2)$ with the ensemble average $\braket{\,\cdot\,}$, dyadic product $\otimes$, translational and rotational diffusion coefficients $D_{\mathrm{T}}$ and $D_{\mathrm{R}}$, respectively, and $2\times 2$-dimensional identity matrix $\mathds{1}_2$.
The self-propulsion of the $i$-th particle is taken into account by the term $v_0\hat{u}(\phi_i)$, where $v_0$ is the propulsion speed of a noninteracting particle and $\hat{u}(\phi_i) = (\cos(\phi_i), \sin(\phi_i))^{\mathrm{T}}$ a unit vector denoting the orientation of the $i$-th particle.  
Furthermore, $\beta =1/(k_{\mathrm{B}}T)$ is the thermodynamic beta with the Boltzmann constant $k_\mathrm{B}$ and absolute temperature $T$ of the particles' environment.
Finally, $\vec{F}_{\mathrm{int}, i}(\{\vec{r}_i\})$ is the interaction force acting on the $i$-th particle. 
It is usually assumed that this force originates from a pair-interaction potential 
$U_2(\norm{\vec{r}_i-\vec{r}_j})$ describing the particle interactions and that the force can be written as   $\vec{F}_{\mathrm{int}, i}(\{\vec{r}_i\}) = -\sum_{j=1, j\neq i}^{N}\Nabla_{\vec{r}_i} U_2(\norm{\vec{r}_i-\vec{r}_j})$ with the nabla operator 
$\Nabla_{\vec{r}_i} = (\partial_{x_{1, i}}, \partial_{x_{2, i}})^\mathrm{T}$ and Cartesian coordinates $x_{1, i}=(\vec{r}_i)_1$ and $x_{2, i}=(\vec{r}_i)_2$.

To derive a field theory for the ABPs from their Langevin equations \eqref{eqn:LangevinR} and \eqref{eqn:LangevinPHI}, we follow a procedure that can be seen as a further development of the derivation presented in Ref.\ \cite{RW}. The main advancements of the new procedure are an adequate consideration of the pair-distribution function and an untruncated consideration of orientational order-parameter fields (see below).    
We start the derivation by rewriting the Langevin equations \eqref{eqn:LangevinR} and \eqref{eqn:LangevinPHI} as the statistically equivalent Smoluchowski equation  
\begin{equation} 
\begin{split}
\dot{P} &= \sum_{i=1}^N \big( (D_\mathrm{T}\Laplace_{\vec{r}_i} + D_\mathrm{R}\partial_{\phi_i}^2 )P
-\Nabla_{\vec{r}_i}\cdot ( v_0 \hat{u}(\phi_i) P )\\
&\qquad\;\;\,\:\! -\Nabla_{\vec{r}_i}\cdot ( \beta D_\mathrm{T} \vec{F}_{\mathrm{int}, i} P) \big),
\end{split}
\end{equation}
which describes the time evolution of the many-particle probability density $P(\{\vec{r}_i\},\{\phi_i\},t)$. Here, the symbol $\Laplace_{\vec{r}_i}=\partial_{x_{1, i}}^2+\partial_{x_{2, i}}^2$ denotes the Laplacian corresponding to $\vec{r}_i$. Integrating both sides of the Smoluchowski equation over all degrees of freedom except for those of one particle, renaming its coordinates as $\vec{r}$ and $\phi$, and multiplying by the particle number $N$, we obtain an equation for the time evolution of the orientation-resolved one-particle density
\begin{equation}
\varrho(\vec{r}, \phi, t) = N \bigg(\prod_{\begin{subarray}{c}j = 1\\j\neq i\end{subarray}}^{N} \int_{\mathbb{R}^2}\!\!\!\!\dif^{2}r_j\int_{0}^{2\pi}\!\!\!\!\!\!\!\dif\phi_j \bigg) P \bigg\rvert_{\begin{subarray}{l}\vec{r}_i=\vec{r},\\\phi_i=\phi\end{subarray}}.
\label{eqn:OPD}%
\end{equation}
By using the divergence theorem and neglecting boundary terms, the equation of motion can be written as
\begin{equation}
\dot{\varrho} = ( D_\mathrm{T}\Laplace_{\vec{r}}+D_\mathrm{R}\partial^2_\phi- v_0\nabla_{\vec{r}}\cdot\hat{u}(\phi) ) \varrho 
+ \mathcal{I}_\mathrm{int}
\label{eq:varrhodot}%
\end{equation}
with the interaction term
\begin{equation}
\begin{split}
\mathcal{I}_\mathrm{int} &= \beta D_T\Nabla_{\vec{r}}\cdot \bigg( \varrho(\vec{r}, \phi, t)\int_{\mathbb{R}^2}\!\!\!\!\dif^2r'\, U_2'(\norm{\vec{r}-\vec{r}'})\\
&\quad\ \frac{\vec{r}-\vec{r}'}{\norm{\vec{r} - \vec{r}'}} \int_{0}^{2\pi}\!\!\!\!\!\!\!\dif\phi'\, g(\vec{r}, \vec{r}', \phi, \phi', t)
\varrho(\vec{r}', \phi', t) \bigg).
\end{split}
\label{eq:Iint}%
\end{equation}
Here, we used the shorthand notation $U_2'(r)\equiv\dif U_2(r)/\dif r$. The pair-distribution function $g(\vec{r}, \vec{r}', \phi, \phi', t)$ gives the relation between the two-particle density $\varrho^{(2)}(\vec{r}, \vec{r}', \phi, \phi', t)$ and the one-particle density:
\begin{equation}
\varrho^{(2)}(\vec{r}, \vec{r}', \phi, \phi', t) = g(\vec{r}, \vec{r}', \phi, \phi', t)\varrho(\vec{r}, \phi, t)\varrho(\vec{r}', \phi', t).
\end{equation}

\begin{figure}[htb]
\centering
\includegraphics[width=8.64cm]{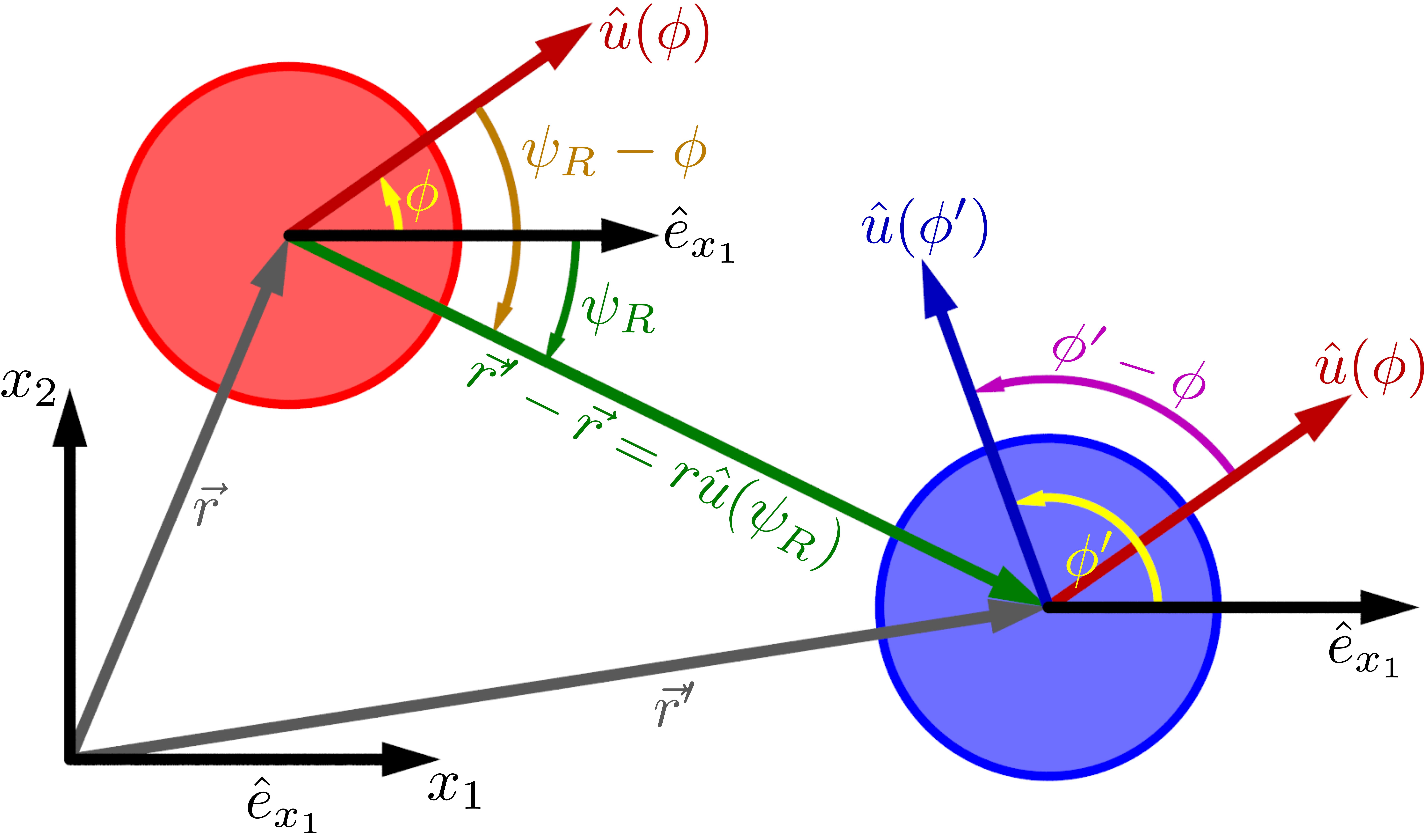}%
\caption{\label{fig:geometry}Absolute and relative positions and orientations of two active Brownian spheres. The unit vector in $x_1$ direction is denoted by $\hat{e}_{x_1}$.}
\end{figure}

As an approximation, we assume that the pair-distribution function can be replaced by that for a corresponding homogeneous and stationary system. 
Using its translational, rotational, and temporal invariance, we can substitute $g(\vec{r}, \vec{r}', \phi, \phi', t)$ by $g(r, \psi_R-\phi, \phi'-\phi)$ with the distance $r$ and the angle $\psi_R$ defined by the parametrization $\vec{r}'-\vec{r}=r\hat{u}(\psi_R)$ (see Fig.\ \ref{fig:geometry}). 
Note that for the system considered here, the latter pair-distribution function has the symmetry property
\begin{equation}
g(r,\psi_R-\phi, \phi'-\phi) = g(r, \phi-\psi_R, \phi-\phi').
\label{eqn:hiddSymm}%
\end{equation}  
In Ref.\ \cite{RW}, the pair-distribution function is further simplified by neglecting its dependence on $\phi'-\phi$, but we omit this additional approximation here. 
Instead, we represent the pair-distribution function by an exact Fourier expansion in the angles $\theta_1 = \psi_R-\phi$ and $\theta_2 = \phi' -\phi$. Taking the symmetry property \eqref{eqn:hiddSymm} and the fact that $g$ is real into account, the expansion reads
\begin{equation}
g(r, \theta_1, \theta_2) = \sum_{k_1,k_2 = -\infty}^{\infty} g_{k_1 k_2}(r)\cos(k_1\theta_1+k_2\theta_2)
\label{eqn:2Fexp}%
\end{equation}
with the $r$-dependent expansion coefficients 
\begin{equation}
g_{k_1 k_2}(r) = \frac{\TINT{0}{2\pi}{\theta_1}\TINT{0}{2\pi}{\theta_2}\, g(r, \theta_1, \theta_2)\cos(k_1\theta_1+k_2\theta_2)}{\pi^2(1+\Kronecker{k_1}{0})(1+\Kronecker{k_2}{0})}.
\end{equation}
In Ref.\ \cite{RW}, this expansion is carried out only up to second order in $\theta_1$ and zeroth order in $\theta_2$, neglecting higher-order contributions. 

Analogously to the orientational expansion of the pair-distribution function into a Fourier series, we perform an orientational expansion of the one-particle density $\varrho(\vec{r}, \phi, t)$ into Cartesian tensor order-parameter fields $\mathcal{O}_{i_1\dotsb i_n}(\vec{r}, t)$ \cite{GubbinsGray,teVrugt19}.
This exact orthogonal expansion is given by\footnote{From here on, summation over indices appearing twice in a term is implied.}  
\begin{equation}
\varrho(\vec{r}, \phi, t) = \sum_{n= 0}^{\infty} \mathcal{O}_{i_1\dotsb i_n}(\vec{r}, t) 
u_{i_1}(\phi)\dotsb u_{i_n}(\phi)
\label{eqn:ODEx2}%
\end{equation}
with the Cartesian tensor order-parameter fields
\begin{equation}
\mathcal{O}_{i_1\dotsb i_n}(\vec{r}, t) = \int_{0}^{2\pi}\!\!\!\!\!\!\!\dif\phi\, \mathfrak{U}_{i_1\dotsb i_n}(\phi)\varrho(\vec{r}, \phi, t) 
\label{eqn:ODEx}%
\end{equation}
and the unit-vector elements $u_{i}=(\hat{u})_i$. 
The orientation-dependent tensors $\mathfrak{U}_{i_1\dotsb i_n}(\phi)$ are defined by the expansion \eqref{eqn:ODEx2} and given in Ref.\ \cite{teVrugt19}.  
Additionally, we perform the angular integration in Eq.\ \eqref{eq:Iint} and an untruncated gradient expansion \cite{Yang1976,Evans1979,EmmerichLWGTTG12} to remove the spatial convolution integral. 
In the resulting equation, the contributions corresponding to the time evolutions of the individual order-parameter fields \eqref{eqn:ODEx} can be separated by making use of the orthogonality of the expansion \eqref{eqn:ODEx2}.     

The described procedure yields the dynamic field equations  
\begin{widetext}
\begin{equation}
\begin{split}%
\dot{\mathcal{O}}_{i_1\dotsb i_d} =& \sum_{k=0}^{\infty}\int_{0}^{2\pi}\!\!\!\!\!\!\!\dif\phi\,  \mathfrak{U}_{i_1\dotsb i_d}\bigg(
(D_\mathrm{T}\Laplace_{\vec{r}} + D_\mathrm{R} \partial^2_\phi - v_0\partial_i u_i(\phi))u_{j_1}(\phi)\dotsb u_{j_k}(\phi)\mathcal{O}_{j_1\dotsb j_k}(\vec{r}, t)\\
&+\partial_{l_0}\bigg(\sum_{h=0}^{\infty}u_{f_1}(\phi)\dotsb u_{f_h}(\phi)\mathcal{O}_{f_1\dotsb f_h}(\vec{r}, t)\sum_{m = 0}^{\infty}\frac{1}{m!} \sum_{k_1, k_2=-\infty}^{\infty}\frac{1}{\pi^2}A(m, k_1,k_2)\\
&\qquad\quad\, M_{abc}u_a((k_1+k_2)\phi) \mathcal{C}^{(k_1)}_{b,l_0l_1 \dotsb l_m} \nabla_{l_1\dotsb l_m} \sum_{p = 0}^{\infty} \mathcal{C}^{(k_2)}_{c,n_1\dotsb n_p}\mathcal{O}_{n_1\dotsb n_p}(\vec{r}, t)
\bigg)\!\bigg).
\end{split}%
\label{eqn:formalDynEqOP}%
\end{equation}
\end{widetext}
In these field equations, the tensor 
$M_{ijk} = \Kronecker{i}{j}\Kronecker{k}{1} + \Kronecker{i}{k}\Kronecker{j}{1} - \Kronecker{j}{k}\Kronecker{i}{1}$
as well as the operator 
$\nabla_{i_1 \dotsb i_m} = \partial_{i_1}\!\dotsb \partial_{i_m}$ 
with $\partial_i\equiv\partial/\partial_{x_i}$ and $x_i=(\vec{r})_i$ 
are introduced to make the notation more compact.
Furthermore, Eq.\ \eqref{eqn:formalDynEqOP} contains the radial coefficients
\begin{equation}
A(m, k_1,k_2) = -\pi^2\beta D_\mathrm{T}\int_{0}^{\infty}\!\!\!\!\!\!\!\:\!\dif r\,  r^{m+1}U_2'(r)g_{k_1 k_2}(r)
\label{eq:Amk1k2}%
\end{equation}
with the symmetry property $A(m, k_1, k_2) = A(m, -k_1,-\!k_2)$ and
the circular coefficient tensors
\begin{equation}
\mathcal{C}^{(k)}_{i,j_1 \dotsb j_m} = \int_{0}^{2\pi}\!\!\!\!\!\!\!\dif\phi\, u_i(k\phi) u_{j_1}(\phi)\dotsb u_{j_m}(\phi)
\end{equation}
with the property 
$\mathcal{C}^{(k)}_{i,j_1\dotsb j_m} = 0\;\; \forall\, k>m$.
Equation \eqref{eqn:formalDynEqOP} describes the time evolution of the orientational order-parameter fields \eqref{eqn:ODEx} and constitutes our general field theory, which is the central result of this work.

To calculate specific values of the coefficients \eqref{eq:Amk1k2}, the pair-interaction potential $U_2(r)$ must be specified and sufficient knowledge about the functions $g_{k_1 k_2}(r)$ is needed as input. It is not necessary to know the full functions $g_{k_1 k_2}(r)$. Since they appear only in a product with the interaction force $-U_2'(r)$, it is sufficient to know the course of these functions for all $r$ where $-U_2'(r)$ is considerably large.
In the existing simulation studies on systems of active Brownian spheres, a Weeks-Chandler-Andersen potential is usually considered. For this choice of $U_2(r)$, an analytic representation of the function $-U_2'(r)g(r,\theta_1,\theta_2)$ can be found in Ref.\ \cite{Jeggle2019}. We use this representation in section \ref{ssec:MIPS} further below.

\subsection{\label{ssec:APPROXIMATIONS}Approximations}
The general field equations \eqref{eqn:formalDynEqOP} contain order-parameter fields (see indices $k$, $h$, and $p$) and derivatives (see index $m$) up to infinite order. To obtain a finite model that is easier to analyze and apply, Eqs.\ \eqref{eqn:formalDynEqOP} can be approximated by truncating the summation over these indices at the desired orders and thus \textit{limiting the maximal order of the order-parameter fields and derivatives}. In this way, the field theory \eqref{eqn:formalDynEqOP} constitutes a general framework that provides various special models for particular applications. 
Usually, one would truncate the order-parameter fields at zeroth to second order and the derivatives at second to sixth order.

The first three of the orientational order-parameter fields \eqref{eqn:ODEx} are well known from liquid crystal theory \cite{deGennesP1995}. They are the scalar density field $\rho(\vec{r}, t)$, which describes the local number density of the particles,\footnote{The actual particle number density or number concentration of the ABPs is given by $2\pi\rho(\vec{r}, t)$.} the polarization vector field $\mathcal{P}_i(\vec{r}, t)$, which describes the local mean particle orientation and amount of alignment, and the symmetric and traceless nematic tensor field $\mathcal{Q}_{ij}(\vec{r}, t)$, which describes the preferred orientation and amount of local parallel or anti-parallel alignment. As is usual in local field theories for liquid crystals, we restrict our orientational expansion to these order-parameter fields in the following. 
This implies a second-order approximation of the one-particle density: \begin{equation}
\varrho(\vec{r}, \phi ,t) \approx \rho(\vec{r}, t) 
+ \mathcal{P}_i(\vec{r}, t) u_i(\phi)
+ \mathcal{Q}_{ij}(\vec{r}, t) u_i(\phi)u_j(\phi) .
\end{equation}
The order-parameter fields $\rho$, $\mathcal{P}_i$, and $\mathcal{Q}_{ij}$ can be obtained from the one-particle density $\varrho$ by
\begin{align}
\rho(\vec{r}, t) &= \frac{1}{2\pi} \int_{0}^{2\pi}\!\!\!\!\!\!\dif\phi\, \varrho(\vec{r}, \phi ,t),
\label{eqn:projRho}\\
\mathcal{P}_i(\vec{r}, t) &= \frac{1}{\pi} \int_{0}^{2\pi}\!\!\!\!\!\!\dif \phi\, u_i\varrho(\vec{r}, \phi ,t),
\label{eqn:projP}\\
\mathcal{Q}_{ij}(\vec{r}, t) &= \frac{2}{\pi} \int_{0}^{2\pi}\!\!\!\!\!\!\dif \phi\, \Big(u_iu_j-\frac{1}{2}\Kronecker{i}{j}\Big)\varrho(\vec{r}, \phi ,t),
\label{eqn:projQ}%
\end{align}
which allows to identify the tensors $\mathfrak{U}_{i_1\dotsb i_n}(\phi)$ corresponding to the chosen order-parameter fields. 

When truncating the derivatives at second order, the resulting model is able to describe the enhanced mobility of the ABPs and the onset of instabilities. This is sufficient to obtain, e.g., the spinodal for MIPS \cite{Bialk2013,RW, Nie2019}. To describe not only the onset of structure formation like MIPS, but also the emerging patterns and their time evolution, a model of fourth order in derivatives is required \cite{Speck2014,Wittkowski2014,Tjhung2018}. 
When the model includes derivatives up to sixth order, it is even able to describe crystallization of ABPs and the particle-resolved lattice structures \cite{Menzel2013,Menzel2014, Alaimo2016,Alaimo2018,Praetorius2018}. It is reasonable to truncate the derivatives at an even order, since otherwise terms that include only $\partial_i$ and $\rho$ but not $\mathcal{P}_i$ or $\mathcal{Q}_{ij}$ could not contribute to the dynamic equation of $\rho$ at the highest considered order in derivatives.

Reducing the general field theory \eqref{eqn:formalDynEqOP} to equations that are of second order in the order-parameter fields and of sixth order in the derivatives leads to a still rather complicated model that is accompanied by a number of exceptionally long equations. Therefore, we further simplify the equations by performing an additional approximation that is known as \textit{quasi-stationary approximation} (QSA) \cite{Cates2013,RW}. This approximation makes use of the fact that the density $\rho$ is a conserved quantity, whereas the order parameters $\mathcal{P}_i$ and $\mathcal{Q}_{ij}$ are not. Since the relaxation time of a conserved quantity is typically much larger than that of a nonconserved one, the dynamics described by the model is considered on the typical time scale of the density $\rho$, where $\mathcal{P}_i$ and $\mathcal{Q}_{ij}$ can be considered as relaxing instantaneously. This leads to constitutive equations for $\mathcal{P}_i$ and $\mathcal{Q}_{ij}$. Keeping the maximal order in derivatives of the equations for $\rho$, $\mathcal{P}_i$, and $\mathcal{Q}_{ij}$ fixed, a recursive application of the constitutive equations results in a dynamic equation for $\rho$ and explicit equations for $\mathcal{P}_i$ and $\mathcal{Q}_{ij}$ that involve only $\rho$ and its derivatives. This procedure (see Refs.\ \cite{Cates2013,RW} for details) thus reduces the initially three coupled and time-dependent partial differential equations for $\rho$, $\mathcal{P}_i$, and $\mathcal{Q}_{ij}$ with these three order-parameter fields as unknown functions to only one partial differential equation for $\rho$ with $\rho$ as the only remaining unknown function. In this way, the complexity of the model is strongly reduced. 

When the model obtained by these approximations still involves too many terms, one can consider the combined order in the derivatives $\partial_i$ and density $\rho$ of each term and discard all terms whose combined order exceeds a certain maximum value. In the present work, we define the combined order as the sum of the order in $\partial_i$ and the order in $\rho$, but in principle one could also assign different weights to $\partial_i$ and $\rho$. Using our way of counting orders, the truncation of the combined order constitutes a \textit{low-density approximation}.   

We stress that the approximations described in this section are not necessary. They could be weakened or completely omitted when a more accurate and complex model is wanted.

\section{\label{sec:MODELS}Special cases and comparison with other field theories}
In this section, we present and discuss three special models obtained by applying the aforementioned approximations to Eq.\ \eqref{eqn:formalDynEqOP}. For each model, we included the order-parameter fields $\rho$, $\mathcal{P}_i$, and $\mathcal{Q}_{ij}$ and performed a QSA. In the first and second model, we considered derivatives up to second and fourth order, respectively, without an additional limitation of the combined order. As a full model of sixth order in derivatives would be too complicated to be presented here, we show a model with a maximal combined order of seven as a third special model. 
The models consist of the continuity equation
\begin{equation}
\dot{\rho} = \partial_i J_i
\label{eqn:ConsservativeForm}%
\end{equation}
for the conserved density $\rho$, where $J_i$ is a model-dependent current, and explicit constitutive equations for the nonconserved polarization $\mathcal{P}_i$ and nematic tensor $\mathcal{Q}_{ij}$.    
While we present the equations for $\rho$ completely, in the equations for $\mathcal{P}_i$ and $\mathcal{Q}_{ij}$ terms of the highest one and two orders in derivatives, respectively, that are considered in the equation for $\rho$ are not shown. The reason for this is that within the QSA these terms do not contribute to the given equation for $\rho$, since in the scalar continuity equation $\mathcal{P}_i$ and $\mathcal{Q}_{ij}$ are always accompanied by one and two derivatives, respectively.

We compare our special models with various popular models from the literature and show that, when considering a one-component system of ABPs in two spatial dimensions, those models can be identified as limiting cases of ours.\footnote{Some models from the literature consider more general systems that include, e.g., mixtures of different types of ABPs, run-and-tumble motion, and three spatial dimensions. These models arise as limiting cases of our models not in general, but when focusing on the ABP system considered in the present article.} Not included in the comparison are the models from Refs.\ \cite{Steffenoni17,Grossmann19}, since they consider systems that are inherently different from the one the present article is based on. In the work \cite{Steffenoni17}, the flow field of a ABP suspension is explicitly considered, and in the work \cite{Grossmann19}, an explicit particle-field representation is used over which, e.g., the particle interactions are defined.

\subsection{\label{sec:2ndorderGradientModel}$\boldsymbol{2}$nd-order-derivatives model}
Our first special model, which contains derivatives up to second order, is given by the density current
\begin{equation}
J_i = D(\rho)\partial_i\rho
\label{eqn:Dyn2DwD}%
\end{equation}
with the density-dependent diffusion coefficient 
\begin{equation}
\begin{split}
D(\rho) &= D_\mathrm{T} +2A(1, 0,0)\rho \\
&\quad\, +\frac{1}{2D_R}\left(v_0  -4(A(0, 1,0)+A(0, 1,-\!1))\rho\right)\\
&\qquad\qquad \ \;  \left(v_0 -8A(0, 1,0)\rho\right)
\end{split}
\label{eqn:diffcoefficient}%
\end{equation}
and the constitutive equations 
\begin{equation}
\mathcal{P}_i = -\frac{v_0-8A(0, 1,0)\rho}{D_\mathrm{R}}\partial_i\rho 
\label{eq:model2P}%
\end{equation}
and
\begin{equation}
\mathcal{Q}_{ij} = 0.
\label{eq:model2Q}%
\end{equation}
It is the model with the lowest nontrivial order in derivatives that constitutes a special case of our general theory. 

Other models of this order in derivatives have previously been proposed in Refs.\ \cite{Bialk2013,Cates2013,RW}. 
The model given by Eqs.\ (19) and (20) in Ref.\ \cite{Bialk2013} is obtained from our Eqs.\ \eqref{eqn:ConsservativeForm}-\eqref{eq:model2P} when we neglect the coefficient $A(1,0,0)$, which originates from the gradient expansion, as well as the coefficient $A(0, 1,-1)$, which is related to the last argument of the pair-distribution function $g(r, \psi_R-\phi, \phi'-\phi)$ and vanishes when that dependence is ignored.  
The parameter $\zeta$ in the model of Ref.\ \cite{Bialk2013} can be related to our coefficients by $\zeta = 4A(0, 1,0)$. 
When we consider the model given by Eqs.\ (7)-(9) in Ref.\ \cite{Cates2013} and set the number of spatial dimensions to $d=2$ as well as the run-and-tumble rate to $\alpha=0$ so that the system studied in Ref.\ \cite{Cates2013} corresponds to our ABP system, their model becomes similar to ours. Equivalence of both models is reached when we identify the phenomenological propulsion speed $v(\rho)$ occurring in their model as $v(\rho)=v_0-4A(0,1,0)\rho$ (see section \ref{sec:v} for details).  
In the model given by Eqs.\ (15), (16), (B16), and (B17) in Ref.\ \cite{RW}, we consider the one-component case (see Eq.\ (51) in Ref.\ \cite{RW}). A comparison with our model shows that their model follows from ours when we neglect the coefficient $A(0,1,-1)$ and that their coefficients are related to ours by $a^{(AA)}_0 = A(0, 1,0)/6$ and $a_1^{(AA)} = 4 A(1, 0,0)$. 

Note that the notation of the present work in similar to that of Ref.\ \cite{Cates2013}, but slightly different from that of Refs.\ \cite{Bialk2013,RW}. In the latter references, the density field is defined with an additional factor of $2\pi$ so that it is equivalent to $2\pi\rho$ in our notation. Furthermore, Ref.\ \cite{Bialk2013} defines the polarization field with an additional factor of $\pi$ so that it is equivalent to $\pi\mathcal{P}_i$ in our notation.

\subsection{$\boldsymbol{4}$th-order-derivatives model}
The second special model contains derivatives up to fourth order and is given by the density current 
\begin{equation}
\begin{split}
J_i &= (\alpha_{1}+\alpha_{2}\rho+\alpha_{3}\rho^2)\partial_i\rho \\
&\quad +(\alpha_{4}+\alpha_{5}\rho+\alpha_{6}\rho^2+\alpha_{7}\rho^3+\alpha_{8}\rho^4)\partial_i\Laplace\rho\\
&\quad +(\alpha_{9}+\alpha_{10}\rho+\alpha_{11}\rho^2+\alpha_{12}\rho^3)(\Laplace\rho)\partial_i\rho\\
&\quad +(\alpha_{13}+\alpha_{14}\rho+\alpha_{15}\rho^2+\alpha_{16}\rho^3)(\partial_j\rho)\partial_i\partial_j\rho\\
&\quad +(\alpha_{17}+\alpha_{18}\rho+\alpha_{19}\rho^2)(\partial_j\rho)(\partial_j\rho)\partial_i\rho
\end{split}
\label{eqn:4thorderCurrentJ}%
\end{equation}
and the constitutive equations 
\begin{equation}
\begin{split}
\mathcal{P}_i &= (\beta_{1}+\beta_{2}\rho)\partial_i\rho\\
&\quad +(\beta_{3}+\beta_{4}\rho+\beta_{5}\rho^2+\beta_{6}\rho^3)\partial_i\Laplace\rho\\
&\quad +(\beta_{7}+\beta_{8}\rho+\beta_{9}\rho^2)(\Laplace\rho)\partial_i\rho\\
&\quad +(\beta_{10}+\beta_{11}\rho+\beta_{12}\rho^2)(\partial_j\rho)\partial_i\partial_j\rho\\
&\quad +(\beta_{13}+\beta_{14}\rho)(\partial_j\rho)(\partial_j\rho)\partial_i\rho
\end{split}%
\label{eqn:4thorderP}%
\end{equation}
and
\begin{equation}
\begin{split}
\mathcal{Q}_{ij} &= (\gamma_1+\gamma_2\rho+\gamma_3\rho^2)(2\partial_i\partial_j\rho+\Kronecker{i}{j}\Laplace\rho)\\
&\quad +(\gamma_2+2\gamma_3\rho)(-2(\partial_i\rho)\partial_j\rho+\Kronecker{i}{j}(\partial_k\rho)\partial_k\rho),
\end{split}
\label{eqn:4thorderQ}%
\end{equation}
where explicit expressions for the coefficients $\alpha_1,\dotsc,\alpha_{19}$, $\beta_1,\dotsc,\beta_{14}$, as well as $\gamma_1$, $\gamma_2$, and $\gamma_3$ are given in the Appendix. This $4$th-order-derivatives model is the first model presented here, where the nematic tensor $\mathcal{Q}_{ij}$ contributes to the dynamics of the density $\rho$. 
Setting all coefficients but $\alpha_{1}, \alpha_{2}, \alpha_{3}, \beta_{1}$, and $\beta_{2}$ to zero yields the $2$nd-order-derivatives model \eqref{eqn:ConsservativeForm}-\eqref{eq:model2Q} presented in the previous section. 

We compare the $4$th-order-derivatives model with the models proposed in Refs.\ \cite{Stenhammar2013,Wittkowski2014,Tjhung2018}, which provide an equation only for the density field. 
In the phenomenological model given by Eqs.\ (10)-(13) in Ref.\ \cite{Stenhammar2013}, we neglect the term $\mu_{\mathrm{rep}}$, which was originally inserted into the model to mimic excluded-volume interactions, and the stochastic term with the noise vector $\vec{\Lambda}$, since both of these contributions cannot exist in our predictive deterministic model. The term $\mu_{\mathrm{rep}}$ yields a contribution proportional to $\rho^5\partial_i\rho$ to the current $J_i$ being incompatible with our gradient expansion of the interaction term for a homogeneous system, which gives only terms where the order in $\rho$ is up to one higher than the order in $\partial_i$.
After the two neglections, the model from Ref.\ \cite{Stenhammar2013} can be identified as a limiting case of our $4$th-order-derivatives model. 
Their model is then obtained from ours, when we set $\alpha_4,\alpha_9,\alpha_{13},\dots,\alpha_{19}=0$, $\beta_1,\dotsc,\beta_{14}=0$, and $\gamma_1,\gamma_2,\gamma_3=0$, assume the relations stated in table \ref{tab:CoefficientComparison37} between their coefficients and ours, and perform a nondimensionalization of our model.
\begin{table}[htb]
\caption{\label{tab:CoefficientComparison37}Relations of the coefficients from the model given by Eqs.\ (10)-(13) in Ref.\ \cite{Stenhammar2013} and the $4$th-order-derivatives model given by Eqs.\ \eqref{eqn:ConsservativeForm} and \eqref{eqn:4thorderCurrentJ}-\eqref{eqn:4thorderQ} in the present work. 
The symbols $k_0$ and $\rho_0$ denote parameters of the model from Ref.\ \cite{Stenhammar2013}, the characteristic length $q_0$ and the characteristic time $t_0$ stem from the nondimensionalization underlying this model, and $\alpha_i$, $\beta_i$, and $\gamma_i$ are coefficients of our model.}
\renewcommand{\arraystretch}{1.5}%
\begin{tabular}{|c|c|}
\hline
Relations of the coefficients & \begin{tabular}[c]{@{}c@{}}Corresp.\ term in the\\[-5pt]model from Ref.\ \cite{Stenhammar2013}\end{tabular} \\
\hline
$1=\frac{q_0^2}{t_0}\alpha_1 = -\rho_{0} \frac{q_0^2}{3t_0} \alpha_2 = \rho_{0}^2\frac{q_0^2}{2t_0}\alpha_3$ & $\partial_i\rho$\\
\begin{tabular}[c]{@{}l@{}}$k_0 = -\rho_{0}\frac{q_0^4}{t_0} \alpha_5 = \rho_{0}^2 \frac{q_0^4}{3t_0} \alpha_6 = -\rho_{0}^3 \frac{q_0^4}{3t_0} \alpha_7$ \\ \hskip2.9ex $=\rho_{0}^4\frac{q_0^4}{t_0} \alpha_8$\end{tabular} & $\rho\partial_i\Laplace\rho$\\
$k_0 = \rho_{0} \frac{q_0^4}{t_0} \alpha_{10} = -\rho_{0}^2 \frac{q_0^4}{2t_0} \alpha_{11} = \rho_{0}^3 \frac{q_0^4}{t_0} \alpha_{12} $ & $(\Laplace\rho)\partial_i\rho$\\
$\alpha_4,\alpha_9,\alpha_{13},\dots,\alpha_{19}=0$ & -- \\
$\beta_1,\dotsc,\beta_{14}=0$ & -- \\
$\gamma_1,\gamma_2,\gamma_3=0$ & -- \\[1pt]
\hline
\end{tabular}
\end{table}

Also the models AMB and AMB+ from Refs.\ \cite{Wittkowski2014,Tjhung2018} can be identified as limiting cases of our $4$th-order-derivatives model.
AMB+, which is given by Eqs.\ (3), (5), and (6) in Ref.\ \cite{Tjhung2018}, can be obtained from our $4$th-order-derivatives model by setting  
$\alpha_{2},\alpha_{5},\alpha_{7},\alpha_{8},\alpha_{10},\alpha_{11},\alpha_{12},\alpha_{14},\dotsc,\alpha_{19}=0$, $\beta_1,\dotsc,\beta_{14}=0$, and $\gamma_1,\gamma_2,\gamma_3=0$, assuming the relations given in table \ref{tab:CoefficientComparison}, and performing a nondimensionalization.
\begin{table}[htb]
\caption{\label{tab:CoefficientComparison}The same as in table \ref{tab:CoefficientComparison37}, but now for the model AMB+ given by Eqs.\ (3), (5), and (6) in Ref.\ \cite{Tjhung2018} with its parameters $a,b,K_0,K_1,\lambda,\zeta,c_\phi,\overline{\rho}$, and $M$.}
\renewcommand{\arraystretch}{1.5}%
\begin{tabular}{|c|c|}
\hline
\;Relations of the coefficients\; & \begin{tabular}[c]{@{}c@{}}\;Corresp.\ term in the\;\\[-5pt]model from Ref.\ \cite{Tjhung2018}\end{tabular} \\[1.0pt]
\hline
$a = q_0^2\frac{\alpha_1-\overline{\rho}^2 \alpha_2}{t_0 M}$ & $\partial_i\rho$\\
$b = \frac{q_0^2 \alpha_2}{3 t_0 c_\phi^2 M}$& $\partial_i(\rho^3)$\\
$K_0 = -q_0^4\frac{\alpha_4+\overline{\rho}\alpha_5}{t_0 M}$& $\partial_i\Laplace\rho$\\
$K_1 = -\frac{q_0^4 \alpha_5}{2 t_0 c_\phi M}$ & $\rho\partial_i\Laplace\rho$, $(\partial_i\partial_j\rho)\partial_j\rho$\\
$\lambda = q_0^4 \frac{\alpha_{13}-2\alpha_5}{2 t_0 c_\phi M}$& $(\partial_i\partial_j\rho)\partial_j\rho$ \\
$\zeta =\frac{q_0^4 \alpha_9}{t_0 c_\phi M}$ & $(\Laplace\rho)\partial_i\rho$ \\
\begin{tabular}[c]{@{}l@{}}$\alpha_{2},\alpha_{5},\alpha_{7},\alpha_{8},\alpha_{10},\alpha_{11},$ \\[-6pt] $\alpha_{12},\alpha_{14},\dotsc,\alpha_{19}=0$\end{tabular} & -- \\
$\beta_1,\dotsc,\beta_{14}=0$ & -- \\
$\gamma_1,\gamma_2,\gamma_3=0$ & -- \\[1pt]
\hline
\end{tabular}
\end{table}
In the course of this nondimensionalization, a dimensionless density field $\phi = c_\phi(\rho-\overline{\rho})$ is introduced, where the constant $c_\phi$ accounts for the nondimensionality of $\phi$ and $\overline{\rho}$ is a reference density.
In AMB and AMB+, this reference density is chosen to be the mean-field critical-point density \cite{Wittkowski2014,Tjhung2018}. 
The mobility $M$ in AMB+ is considered as a constant in Ref.\ \cite{Tjhung2018} to keep the model relatively simple, but in general it could depend on the density $\rho$ and its derivatives. In the $4$th-order-derivatives model of the present work, in contrast, effectively no fixed mobility is assumed.
The model AMB, which is given by Eqs.\ (1)-(3) in Ref.\ \cite{Wittkowski2014}, constitutes a limiting case of the more general model AMB+. It is obtained from AMB+ and from our $4$th-order-derivatives model for $\zeta=0$ and $\alpha_9=0$, respectively.

\subsection{$\boldsymbol{7}$th-order low-density model}
The third model considers terms up to a combined order of seven. It is given by the density current 
\begin{equation}
\begin{split}
J_i &= (\alpha_{1}+\alpha_{2}\rho+\alpha_{3}\rho^2)\partial_i\rho
+(\alpha_{4}+\alpha_{5}\rho+\alpha_{6}\rho^2)\partial_i\Laplace\rho\\
&\quad+(\alpha_{9}+\alpha_{10}\rho)(\Laplace\rho)\partial_i\rho
+(\alpha_{13}+\alpha_{14}\rho)(\partial_i\partial_j\rho)\partial_j\rho\\
&\quad +\alpha_{17}(\partial_j\rho)(\partial_j\rho)\partial_i\rho +\epsilon_1\Laplace^2\partial_i\rho
\end{split}\raisetag{3ex}%
\label{eqn:7thOrderRho}%
\end{equation}
with the coefficient 
\begin{equation}
\epsilon_1 = \dfrac{v_0^2}{512 D_\mathrm{R}^3}\bigg(16D_\mathrm{T}+\dfrac{v_0^2}{D_\mathrm{R}}\bigg)^2+\dfrac{D_\mathrm{T}v_0^4}{128 D_\mathrm{R}^4}
\end{equation}
and the constitutive equations 
\begin{equation}
\begin{split}
\mathcal{P}_i &= (\beta_{1}+\beta_{2}\rho)\partial_i\rho
+(\beta_{3}+\beta_{4}\rho+\beta_{5}\rho^2)\partial_i\Laplace\rho\\
&\quad +(\beta_{7}+\beta_{8}\rho)(\Laplace\rho)\partial_i\rho
+(\beta_{10}+\beta_{11}\rho)(\partial_i\partial_j\rho)\partial_j\rho\\
&\quad +\beta_{13}(\partial_j\rho)(\partial_j\rho)\partial_i\rho-\epsilon_1\dfrac{2}{v_0}\Laplace^2\partial_i\rho
\end{split}\raisetag{5ex}%
\label{eqn:7thOrderLDP}%
\end{equation}
and 
\begin{equation}
\begin{split}
\mathcal{Q}_{ij} &= (\gamma_1+\gamma_2\rho)(2\partial_i\partial_j\rho+\Kronecker{i}{j}\Laplace\rho)\\
&\quad +\gamma_2(-2(\partial_i\rho)\partial_j\rho+\Kronecker{i}{j}(\partial_k\rho)\partial_k\rho).
\end{split}
\label{eqn:7thOrderLDQ}%
\end{equation}
Equations \eqref{eqn:7thOrderRho}-\eqref{eqn:7thOrderLDQ} are a reduced version of the corresponding equations from the $4$th-order-derivatives model, but with an additional term $\propto\Laplace^2\partial_i\rho$ in Eqs.\ \eqref{eqn:7thOrderRho} and \eqref{eqn:7thOrderLDP}. 

Since the inclusion of terms with derivatives up to sixth order is necessary for describing crystals at a particle-resolving length scale, PFC models have typically a combined order of seven. While PFC models are usually derived for systems of passive particles, there exist some \ZT{active PFC models} that describe two-dimensional crystals of ABPs \cite{Menzel2013,Menzel2014,Alaimo2016,Alaimo2018,Praetorius2018}. We will therefore compare these models, which are given by dynamic equations for a rescaled density $\psi$ and polarization $\mathcal{P}_i^0$, with our $7$th-order low-density model. 

The original active PFC model was first proposed in Ref.\ \cite{Menzel2013}, studied in more detail in Refs.\ \cite{Menzel2014,Ophaus2018}, modified in Ref.\ \cite{Alaimo2016}, and extended in Refs.\ \cite{Alaimo2018,Praetorius2018}. 
Its dynamic equation for $\mathcal{P}_i^0$ contains a Toner-Tu term $\propto \mathcal{P}^2\mathcal{P}_i$ \cite{TonerTu1995}, which, depending on the sign of the term's prefactor, encourages or discourages aligned motion. Such an alignment of the particle motion is, however, not included in the Langevin equations \eqref{eqn:LangevinR} and \eqref{eqn:LangevinPHI} of the present work. Furthermore, this term is neglected in all Refs.\ \cite{Menzel2013,Menzel2014,Alaimo2016,Alaimo2018,Praetorius2018}, either directly \cite{Alaimo2018} or during the respective work \cite{Menzel2013,Menzel2014,Alaimo2016,Praetorius2018}. Therefore, we exclude this term from our following comparison.      
Applying a QSA to the original active PFC model, which is given by Eqs.\ (1)-(4) in Ref.\ \cite{Menzel2013} and by Eqs.\ (12) and (13) in Ref.\ \cite{Menzel2014}, one obtains the dimensionless conservation equation
\begin{equation}
t_0\dot{\psi} = q_0 \partial_i J_i^\psi 
\label{eqn:PFCQSAdynamic}%
\end{equation}
with the dimensionless density current 
\begin{equation}
J_i^\psi = (\delta_1+\delta_2\psi+\delta_3\psi^2)q_0\partial_i\psi+\delta_4q_0^3\Laplace\partial_i\psi+\delta_5q_0^5\Laplace^2\partial_i\psi
\label{eqn:PFCQSAJ}%
\end{equation}
and the constitutive equation for the dimensionless polarization vector
\begin{equation}
\begin{split}
{\mathcal{P}}^0_i &= -\frac{\tau_R \tilde{v}_0}{C_1}q_0\partial_i\psi - \frac{\tau_R^2 \tilde{v}_0}{C_1}q_0^3\partial_i\Laplace\psi - \frac{\tau_R^3 \tilde{v}_0}{C_1}q_0^5\partial_i\Laplace^2\psi.
\end{split}%
\label{eqn:PFCQSAP}%
\end{equation}
Here, $q_0$ is the characteristic length and $t_0$ is the characteristic time used for the nondimensionalization of the model.
The coefficients $\delta_1,\dots,\delta_5$ are given by 
\begin{align}
\delta_1 &= 1+\epsilon+3\overline{\psi}^2+\frac{\tau_R \tilde{v}_0}{C_1},\\
\delta_2 &= 6\overline{\psi}, \\
\delta_3 &= 3, \\
\delta_4 &= 2+\frac{\tau_R^2 \tilde{v}_0^2}{C_1}, \\
\delta_5 &= 1+\frac{\tau_R^3\tilde{v}_0^2}{C_1},
\end{align}
where $\tau_R = 1/\tilde{D}_\mathrm{R}$ is a dimensionless rotational relaxation time, $\tilde{D}_\mathrm{R}, \tilde{v}_0, C_1$, and $\epsilon$ are parameters of the PFC model \cite{Menzel2014}, and an overbar denotes a spatial average.
In the case of the parameters $D_\mathrm{R}$ and $v_0$, which occur with the same symbols but different scalings in the PFC model and our models, we inserted a tilde about the parameters from the PFC model to distinguish them from our corresponding parameters.  
Equations \eqref{eqn:PFCQSAdynamic}-\eqref{eqn:PFCQSAP} have to be rescaled if one wants to obtain predictive relations for their phenomenological parameters. The rescaling rules read \cite{Menzel2014}
\begin{align}
\psi &=c_\psi(\rho-\overline{\rho}),\\
\mathcal{P}_i^0 &= \frac{c_\psi}{\sqrt{2}}\mathcal{P}_i
\end{align}
with the constant $c_\psi = \sqrt{u/(\lambda q_0^4)}$ and the reference density $\overline{\rho}$, where $u$ and $\lambda$ are further parameters of the PFC model \cite{Menzel2014}.

When comparing Eqs.\ \eqref{eqn:PFCQSAJ} and \eqref{eqn:PFCQSAP} with Eqs.\ \eqref{eqn:7thOrderRho}, \eqref{eqn:7thOrderLDP}, and \eqref{eqn:7thOrderLDQ}, we find that the former equations are obtained from our $7$th-order low-density model, when setting $\alpha_5, \alpha_6, \alpha_9, \alpha_{10}, \alpha_{13}, \alpha_{14}, \alpha_{17} = 0$, $\beta_2, \beta_4, \beta_5, \beta_7, \beta_8, \beta_{10}, \beta_{11}, \beta_{13} = 0$, as well as $\gamma_1,\gamma_2=0$ and assuming the relations 
\begin{align}
c_\psi &= \sqrt{\frac{t_0 \alpha_3}{3 q_0^2}}, \\
\tau_R &= \frac{\beta_3}{q_0^2 \beta_1}, \\
\tilde{v}_0 &= \frac{\sqrt{2}\beta_1(q_0^6-t_0\epsilon_1)}{q_0 \beta_3^2}, \\
C_1 &= -\frac{2(q_0^6-t_0 \epsilon_1)}{q_0^2 \beta_1 \beta_3}, \\    
\epsilon &= -1 + \frac{t_0\alpha_1}{q_0^2} +\frac{\beta_1}{\sqrt{2}q_0}.
\end{align}
Note that these relations are not unique, since a comparison of Eqs.\ \eqref{eqn:PFCQSAJ} and \eqref{eqn:PFCQSAP} with Eqs.\ \eqref{eqn:7thOrderRho}, \eqref{eqn:7thOrderLDP}, and \eqref{eqn:7thOrderLDQ} leads to an overdetermined system of equations for $c_\psi, \tau_R, \tilde{v}_0, C_1$, and $\epsilon$. 

In Eqs.\ (2)-(4) of Ref.\ \cite{Alaimo2016}, a phenomenologically modified version of the active PFC model is proposed, which uses additional contributions known from the vacancy PFC model \cite{Chan2009,Berry2011,Robbins2012} to penalize negative values of the order parameter $\psi$ that describes the spatial density variation and thus to support an interpretation of density peaks as individual particles. 
Assuming $\psi\neq 0$ and performing a QSA for this model, we obtain the dimensionless density current 
\begin{equation}
\begin{split}
J_i^{\psi} &= (\delta_1+\delta_2\psi+\delta_3\psi^2)q_0\partial_i\psi+(\delta_4+\lambda_1\psi)q_0^3\Laplace\partial_i\psi\\
&\quad+(\delta_5+\lambda_2\psi)q_0^5\Laplace^2\partial_i\psi
\end{split}\raisetag{4ex}
\label{eqn:VPFCQSAJ}%
\end{equation}
and the constitutive equation for the dimensionless polarization vector
\begin{equation}
\begin{split}
\mathcal{P}^0_i &= -\tilde{v}_0 c_\mathcal{P}q_0\partial_i\psi - \tilde{v}_0  \tilde{\alpha}_2 c_\mathcal{P}^2 q_0^3\Laplace \partial_i \psi - \tilde{v}_0  \tilde{\alpha}_2^2 c_\mathcal{P}^3 q_0^5\Laplace^2\partial_i\psi.
\end{split}%
\label{eqn:VPFCQSAP}%
\end{equation}
The coefficients $\delta_1, \dots, \delta_5$, $\lambda_1$, and $\lambda_2$ are now given by
\begin{align}
\delta_1 &= M_0(r+1),\label{qn:VPFdeltasbegin}\\
\delta_2 &= 12 M_0 H \Theta(-\psi)+\tilde{v}_0^2c_\mathcal{P} ,\\
\delta_3 &= 3 M_0 ,\\
\delta_4 &= 2 M_0, \\
\delta_5 &= M_0, \\
\lambda_1 &= \tilde{v}_0^2 \tilde{\alpha}_2 c_\mathcal{P}^2,\\
\lambda_2 &= \tilde{v}_0^2  \tilde{\alpha}_2^2 c_\mathcal{P}^3 \label{qn:VPFdeltasend}%
\end{align}
with the abbreviating notation 
$c_\mathcal{P}=1/(\tilde{\alpha}_4+\beta \Theta(-\psi))$, the parameters $\tilde{v}_0, \tilde{\alpha}_2, \tilde{\alpha}_4, M_0, r, H$, and $\beta$ of the model from Ref.\ \cite{Alaimo2016}, and the Heaviside function $\Theta(x)$. Again, the characteristic length $q_0$ is associated with the nondimensionalization of the model and a tilde is used to distinguish otherwise similar symbols with different meanings. 
To compare with our $7$th-order low-density model, one has to consider the rescaling rules 
\begin{align}
\psi &= c_\psi (\rho - \overline{\rho}),\\
\mathcal{P}^0_i &= c_0\mathcal{P}_i,
\end{align}
where $c_\psi$ and $\overline{\rho}$ are as before and $c_0$ accounts for the nondimensionality of $\mathcal{P}^0_i$. 

The terms proportional to $\lambda_1$ and $\lambda_2$ in Eq.\ \eqref{eqn:VPFCQSAJ} have no counterparts in Eqs.\ \eqref{eqn:PFCQSAJ} and \eqref{eqn:PFCQSAP}. In our $7$th-order low-density model, only the term proportional to $\lambda_2$ is not present. The reason for the absence of this term is that, according to the way of counting orders in this section, this term is of $8$th order. 
When the low-density model is truncated at $8$th or higher order, a corresponding term is included. 
The Heaviside functions, which occur in the coefficients $\delta_2$, $\lambda_1$, and $\lambda_2$ and aim at penalizing negative values of $\psi$, are incompatible with both the model given by Eqs.\  \eqref{eqn:PFCQSAdynamic}-\eqref{eqn:PFCQSAP} and the $7$th-order low-density model. 
Albeit the model given by Eqs.\ \eqref{eqn:PFCQSAdynamic}, \eqref{eqn:VPFCQSAJ}, and \eqref{eqn:VPFCQSAP} being no special case of the $7$th-order low-density model, relations between the phenomenological parameters of the former model and our predictive coefficients can be established by comparing the prefactors of the terms that occur in both models. 
This gives, among others, the relations
\begin{align}
c_\psi &= \sqrt{\frac{q_0^4 \alpha_3}{3\epsilon_1 + 3q_0^2 \overline{\rho}\alpha_4}}, \\
c_\mathcal{P} &= \sqrt{\frac{3 c_0^4 \beta_1^2 \beta_3^2 (\epsilon_1+q_0^2\overline{\rho} \alpha_5)}{q_0^4 t_0^2 \alpha_3 \alpha_5^2}}, \\
\tilde{v}_0 &= -\frac{t_0 \alpha_5}{c_0 q_0 \beta_3}, \\
\tilde{\alpha}_2 &= \sqrt{\frac{t_0^2\alpha_3\alpha_5^2}{3c_0^4\beta_1^4(\epsilon_1+q_0^2\overline{\rho}\alpha_5)}}, \\    
M_0 &= \frac{t_0(\epsilon_1+ q_0^2 \overline{\rho} \alpha_5)}{q_0^6}, \\
r &= -\frac{\epsilon_1 + q_0^2\overline{\rho}\alpha_5-q_0^4(\alpha_1+\overline{\rho}\alpha_2)}{\epsilon_1+q_0^2\overline{\rho} \alpha_5}, \\
H &= \frac{\beta_3(\alpha_2+\overline{\rho}\alpha_3)-\alpha_5\beta_1}{4\alpha_3\beta_3}\sqrt{\frac{q_0^4\alpha_3}{3\epsilon_1+3q_0^2\overline{\rho}\alpha_5}}\quad \text{for } \psi < 0. \raisetag{2ex}
\end{align}

Equations (2)-(6) in Ref.\ \cite{Alaimo2018} constitute a phenomenological extension of the model from Ref.\ \cite{Alaimo2016} towards mixtures. In the limiting case of a one-component system, as it is considered in the present work, the model from Ref.\ \cite{Alaimo2018} reduces to that from Ref.\ \cite{Alaimo2016}. 
The model given by Eqs.\ (5)-(12) in Ref.\ \cite{Praetorius2018} is an extension of the traditional active PFC model \cite{Menzel2013,Menzel2014} from a planar system towards one on a sphere. It contains special differential operators that are defined on a spherical manifold and reduces to the traditional active PFC model in the limiting case of a vanishing local curvature.

\section{\label{sec:Applications}Applications}
The general field theory and special models derived in section \ref{sec:derivation} can be applied to a large number of problems. In this section, we derive the density-dependent mean swimming speed of the particles and show that its behavior is in good agreement with expectations and previous results from the literature. As a further application, we analyze the $2$nd-order-derivatives model to predict the onset of MIPS.

\subsection{\label{sec:v}Density-dependent swimming speed}
The motion of ABPs is typically slowed down by interactions of the particles so that their mean swimming speed $v$ depends on the local density $\rho$ and is smaller than the bare propulsion speed $v_0$. Examples for interactions that reduce $v$ are steric repulsions \cite{Stenhammar2014} and more complicated interactions via, e.g., quorum sensing \cite{Whiteley2017} and visual perception \cite{Lavergne2019}. 
The dependence of the mean swimming speed $v$ on the density $\rho$ is of considerable interest, since it helps to characterize a system of ABPs. In particular, a sufficiently steep decrease of $v$ for growing $\rho$ is known to indicate the emergence of MIPS \cite{Tailleur2008,Cates2013}.  

The mean swimming speed $v$ can depend also on the derivatives of the density $\rho$. In general, it is a functional $v[\rho]$ of $\rho$. To calculate $v[\rho]$, one can write the right-hand side of Eq.\ \eqref{eq:varrhodot} as the sum of a convective contribution and a diffusive remainder. Writing the convective contribution as $-\Nabla\cdot (v[\rho]\hat{u}(\phi)\varrho(\vec{r},\phi,t))$ then allows to identify the expression for $v[\rho]$.  
Applying to this expression the same approximations and expansions as in section \ref{ssec:gfe} as well as a QSA, to zeroth order in derivatives we obtain the density-dependent swimming speed
\begin{equation}
v(\rho) = v_0-4A(0, 1,0)\rho.
\label{eqn:SwimSpeed}%
\end{equation}
The predicted linear decrease of $v(\rho)$ is in very good agreement with previous results of simulations \cite{Fily2012,Stenhammar2013,Stenhammar2014} and analytic considerations \cite{Fily2012,Bialk2013,Cates2013,Stenhammar2013,Speck2014,Speck2015,Sharma2016, RW} from the literature.
By a comparison of the equations for $v(\rho)$ proposed in Refs.\ \cite{Fily2012,Bialk2013,Cates2013,Speck2014,Speck2015} with Eq.\ \eqref{eqn:SwimSpeed}, their estimated threshold density $\rho_0$ \cite{Cates2013} and phenomenological parameters $\lambda$ \cite{Fily2012} and $\zeta$ \cite{Bialk2013,Speck2014,Speck2015} can be identified as $\rho_0=v_0/(4A(0,1,0))$ and $\lambda,\zeta=4A(0, 1,0)$. 
In Ref.\ \cite{Stenhammar2014,Sharma2016}, a similar linear dependence of $v(\rho)$ on $\rho$ was found for a system of ABPs in three spatial dimensions. 
Reference \cite{Cates2010PNAS} considers a system of reproducing bacteria and uses the function $v(\rho) = v_0 e^{-c_\rho \rho}$ for the density-dependent swimming speed. This is a nonlinear function, but it reduces to Eq.\ \eqref{eqn:SwimSpeed} in the low-density limit, where the constant $c_\rho$ can be identified as $c_\rho=4A(0, 1,0)/v_0$.

\subsection{\label{ssec:MIPS}Predictions for motility-induced phase separation}
To predict the onset of MIPS as a function of the activity and mean density of the ABPs, it is sufficient to consider our $2$nd-order-derivatives model from section \ref{sec:2ndorderGradientModel} and to perform a linear stability analysis. 
For models of this structure, the stability analysis leads to the spinodal condition \cite{Bialk2013,Cates2013,Speck2014,RW}
\begin{equation}
D(\rho) = 0,
\label{eqn:spinCon}%
\end{equation}
where the density-dependent diffusion coefficient $D(\rho)$ is here given by Eq.\ \eqref{eqn:diffcoefficient}. 
This condition contains the three coefficients $A(1,0,0)$, $A(0,1,-1)$, and $A(0,1,0)$ and generalizes previously derived spinodal conditions from the literature. When neglecting the coefficients $A(1,0,0)$ and $A(0,1,-1)$, Eq.\ \eqref{eqn:spinCon} reduces to the spinodal condition from Ref.\ \cite{Bialk2013}, and neglecting only $A(0,1,-1)$ gives the spinodal condition for a one-component system of ABPs from Ref.\ \cite{RW}. 

\begin{figure}[htb]
\centering
\includegraphics[width = \linewidth]{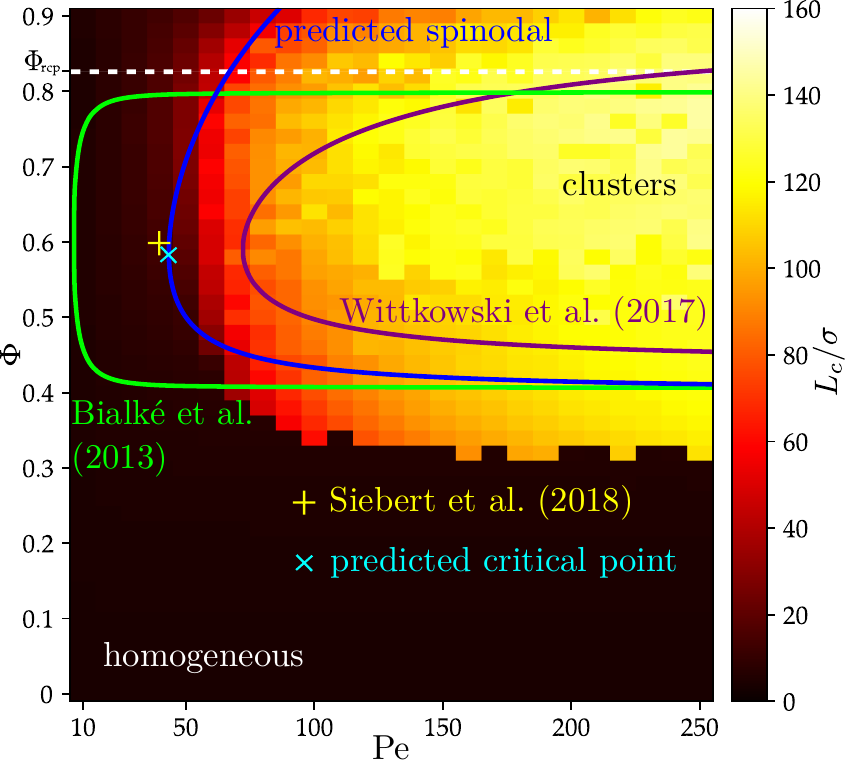}
\caption{\label{fig:MIPS}State diagram showing particle-based simulation data for the characteristic length $L_c$ from Ref.\ \cite{Jeggle2019} and theoretical predictions for the spinodal from Refs.\ \cite{Bialk2013,RW} and from the spinodal condition \eqref{eqn:spinCon} as a function of P\'{e}clet number $\mathrm{Pe}$ and packing density $\Phi$. High $L_c/\sigma$ correspond to clusters whereas low $L_c/\sigma$ mean that the system is in a homogeneous state. An estimate for the critical point from Ref.\ \cite{Siebert2018} and the critical point calculated from Eq.\ \eqref{eqn:spinCon} are shown as well. The random-close-packing density $\Phi_{\mathrm{rcp}} \approx 0.82$ of hard spheres in two spatial dimensions \cite{Berryman1983} is marked.}
\end{figure}

For the remainder of this section, we specify the pair-interaction potential $U_2(r)$ as the purely repulsive  Weeks-Chandler-Anderson potential
\begin{equation}
U_2(r) = \begin{cases}
4\epsilon\left(\left(\frac{\sigma}{r}\right)^{12}-\left(\frac{\sigma}{r}\right)^{6}\right)+\epsilon, &\text{if } r<2^{\frac{1}{6}}\sigma,\\
0, & \text{else,}
\end{cases}\label{eqn:WCA}%
\end{equation}
where $\sigma$ denotes the effective diameter of the particles and $\epsilon$ determines the interaction strength.
This choice for the interaction potential suits well the behavior of ABPs \cite{Buttinoni2013} and is in line with the most studies on ABPs that are related to the present work.
Another advantage of this choice is the fact that for the Weeks-Chandler-Anderson potential an analytic representation of the pair-distribution function of ABPs in two spatial dimensions is available \cite{Jeggle2019}. Using this representation, concrete values for the coefficients $A(1,0,0)$, $A(0,1,-1)$, and $A(0,1,0)$ can be calculated. They are approximately given by
\begin{align}
A(0, 1,0)& = 5.88,\label{eqn:A010coeff}\\
A(0, 1,-1)& = -0.037 - 2.16 \Phi,\label{eqn:A01m1coeff}\\
A(1, 0,0)& = 6.08 + 2.93e^{2.87\Phi}\label{eqn:A100coeff}
\end{align}
as functions of the mean packing density $\Phi = \rho \sigma^2\pi^2/2$ of the ABPs in the system. With these functions and using the relation $D_R = 3D_T/\sigma^2$, which holds for spherical particles, we can plot our prediction \eqref{eqn:spinCon} for the spinodal as a function of the P\'{e}clet number $\mathrm{Pe} = v_0\sigma/D_T$ and the packing density $\Phi$. In Fig.\ \ref{fig:MIPS}, this spinodal condition is shown together with earlier spinodal conditions from Ref.\ \cite{Bialk2013} (with our values for the coefficients) and from Ref.\ \cite{RW} (with their values for the coefficients\footnote{In Ref.\ \cite{RW}, the spinodal condition contains the coefficients $a^{\mathrm{(AA)}}_0 = 0.86$ and $a_1^{\mathrm{(AA)}} = 24.94 + 11.30e^{2.62\Phi}$. When we make use of the results of section \ref{sec:2ndorderGradientModel} and of Eqs.\ \eqref{eqn:A010coeff}-\eqref{eqn:A100coeff}, we find the slightly different values $a^{\mathrm{(AA)}}_0 = A(0, 1,0)/6 = 0.98$ and $a_1^{\mathrm{(AA)}} = 4 A(1, 0,0) = 24.33 + 11.67e^{2.87\Phi}$ for these coefficients.}). 
For comparison, also results of Brownian dynamics simulations for the state diagram of the considered system of ABPs from Ref.\ \cite{Jeggle2019} are shown. 
In these simulations, the P\'{e}clet number $\mathrm{Pe}$ was varied via the diffusion coefficient $D_T = \sigma^2/\tau_{LJ}$ with the Lennard-Jones time scale $\tau_{LJ} = \sigma^2/(\beta D_T \epsilon)$, whereas the bare propulsion speed $v_0$ was kept at $v_0 = 24\sigma/\tau_{LJ}$.
A comparison of the analytic predictions for the spinodal and the actual state diagram shows that our spinodal condition \eqref{eqn:spinCon} is in very good agreement with the simulation results. Interestingly, this agreement is very good even for packing densities $\Phi$ above the random-close-packing density $\Phi_{\mathrm{rcp}} \approx 0.82$ of hard spheres in two spatial dimensions \cite{Berryman1983}. 
Moreover, the agreement is much better than for the earlier spinodal conditions from the literature. 
Figure \ref{fig:MIPS} shows also the critical point ($\mathrm{Pe}_c \approx 41.5, \Phi_c \approx 0.588$) that results from the spinodal condition \eqref{eqn:spinCon} and the estimate ($\mathrm{Pe}_c = 40, \Phi_c = 0.597$) for the critical point recently proposed in Ref.\ \cite{Siebert2018}. Remarkably, these two results are in excellent agreement. The minor difference of both points is within their numerical inaccuracy.

\section{\label{sec:Conclusion}Conclusions}
Using the Smoluchowski framework and an explicit coarse-graining, we derived a highly accurate and predictive local field theory for spherical ABPs in the plane. An important feature of our field theory is its high generality. It allowed to identify various popular models for ABPs from the literature, including AMB \cite{Wittkowski2014} and its recent extension AMB+ \cite{Tjhung2018}, as limiting cases of the general field theory and thus to obtain explicit expressions for the coefficients occurring in these models. Especially for phenomenological models such a linkage of their so far unspecified coefficients to the microscopic parameters of the system constitutes an important progress. Alongside the general field theory, we presented reduced models that are easier to apply. 
To demonstrate specific applications of the field theory, we derived an expression for the density-dependent mean swimming speed of interacting ABPs and an expression describing the spinodal corresponding to the onset of MIPS. In both cases, we found an excellent agreement of our analytical results with corresponding data from simulations and experiments described in the literature. This agreement was in particular better than for other analytical predictions published earlier. 

The general field theory and reduced models presented in this article can be applied to study a lot of further far-from-equilibrium effects of ABPs. For example, a more detailed analysis of the $4$th-order-derivatives model could reveal effects that are described by its terms of high order in the density, which are not included in AMB+ and the $7$th-order low-density model. 
The latter model could be used to study active solidification and crystallization of ABPs more closely. Comparing the large number of articles focusing on fluid states of active matter with the few publications on solid states of active particles, we can expect that in solid states of ABPs there are many fascinating effects still to discover. 
Our reduced models could also be used for investigating active crystals on curved manifolds like a sphere \cite{Praetorius2018}. Since their dynamic equations for the density field include only a scalar order-parameter field, the differential operators in the models could be straightforwardly adapted to a particular curved manifold, whereas this would be much more challenging when vectorial or higher-order tensorial order-parameter fields are involved \cite{Praetorius2018}.  

Furthermore, the reduced models presented here could be extended. The accuracy of these models could be further increased, e.g., by omitting the QSA and using a model with dynamic equations for all considered orientational order-parameter fields. 
Finally, the general field theory could be extended towards systems of higher complexity. 
Important examples that should be addressed in the near future are extensions towards mixtures of active and passive particles \cite{Stenhammar2015,RW,Alaimo2018}, nonspherical ABPs \cite{Wittkowski2011,Wittkowski2012} and active liquid crystals \cite{DeCamp2015,Doostmohammadi2018,Lemma2019}, as well as systems with three spatial dimensions. 
In the last case, one could use the recently obtained analytical representation for the pair-distribution function of ABPs in three spatial dimensions \cite{Broeker2019} to derive the extended field theory, which would be highly useful as it is known that phase transitions in active matter can strongly depend on the system's dimensionality \cite{Stenhammar2014}.

\begin{acknowledgments}
R.W.\ is funded by the Deutsche Forschungsgemeinschaft (DFG, German Research Foundation) -- WI 4170/3-1. 
\end{acknowledgments}

\appendix
\section{\label{sec:Koeff4th}Coefficients for the 4th-order-derivatives model}
In this appendix, explicit expressions for the coefficients occurring in Eqs.\ \eqref{eqn:4thorderCurrentJ}-\eqref{eqn:4thorderQ} are presented. 
To simplify these expressions, we introduce the rotational relaxation time $\tau = 1/D_\mathrm{R}$.

The coefficients in Eq.\ \eqref{eqn:4thorderCurrentJ} are given by
\begin{align}
\begin{split}
\alpha_{1} &=  D_\mathrm{T}+\frac{1}{2}\tau v_0^2,
\end{split}\\
\begin{split}
\alpha_{2} &=  -2\tau v_0(A(0,1,-1)+3A(0,1,0))+2A(1,0,0),
\end{split}\\
\begin{split}
\alpha_{3} &= 16\tau A(0,1,0)(A(0,1,-1)+A(0,1,0)),
\end{split}\\
\begin{split}
\alpha_{4} &= \frac{1}{32} \tau ^2 v_0^2 (16 D_\mathrm{T}+\tau  v_0^2),
\end{split}\\
\begin{split}
\alpha_{5} &= \frac{1}{8}(-16 D_\mathrm{T} \tau ^2 v_0 (A(0,1,-1)+3 A(0,1,0))\\
&\quad\, +\tau  v_0 (\tau  v_0 (-\tau  v_0 (A(0,1,-1) +5 A(0,1,0)\\
&\quad\, +A(0,1,1))+4 A(1,0,-1)+4 A(1,0,1)\\
&\quad\, +A(1,2,-2)+4 A(1,2,-1) +A(1,2,0))\\
&\quad\, -6A(2,1,-1)-6A(2,1,0))+2 A(3,0,0)),
\end{split}\raisetag{9.5ex}\\
\begin{split}
\alpha_{6} &=  \frac{1}{2} \tau(32 D_\mathrm{T} \tau  A(0,1,0) (A(0,1,-1)+A(0,1,0))\\
&\quad\, +\tau ^2 v_0^2 (A(0,1,-1) (4 A(0,1,0)+A(0,1,1))\\
&\quad\, +A(0,1,0) (9 A(0,1,0)+4 A(0,1,1)))\\
&\quad\, -\tau  v_0(A(0,1,1) A(1,2,-2)+A(0,1,-1) \\
&\qquad\;\, (4A(1,0,-1)+4A(1,0,1)+4A(1,2,-1)\\
&\quad\,+A(1,2,0))+A(0,1,0) (3 (4 A(1,0,-1)\\
&\quad\, +4 A(1,0,1)+A(1,2,-2)+4 A(1,2,-1))\\
&\quad\, +2 A(1,2,0)) ) +A(1,2,-2) A(1,2,0)\\
&\quad\,+12 A(0,1,0) A(2,1,-1)+6 (A(0,1,-1)\\
&\quad\, +A(0,1,0)) A(2,1,0)), 
\end{split}\raisetag{20ex}\\
\begin{split}
\alpha_{7} &= -2 \tau ^2 A(0,1,0)(\tau  v_0 (A(0,1,-1) (5 A(0,1,0)\\
&\quad\, +3 A(0,1,1))+A(0,1,0) (7 A(0,1,0)\\
&\quad\,\quad\, (A(1,0,-1)+5 A(0,1,1))-8A(0,1,-1)\\
&\quad\,-8A(0,1,0)+A(1,0,1))-2A(1,2,-2)\\
&\quad\,\quad\,(A(0,1,0)+A(0,1,1))-8A(1,2,-1)\\
&\quad\,\quad\,(A(0,1,-1)+A(0,1,0)) -A(1,2,0)\\
&\quad\,\quad\,(A(0,1,-1)+A(0,1,0)) )), 
\end{split}\raisetag{13ex}\\
\begin{split}
\alpha_{8} &= 16 \tau^3 A(0,1,0)^2 (A(0,1,-1)+A(0,1,0))\\
&\quad\,\;(A(0,1,0)+A(0,1,1)),
\end{split}\\
\begin{split}
\alpha_{9} &= -\frac{1}{4} \tau  v_0 (16 D_\mathrm{T} \tau  A(0,1,0)+2A(2,1,0)\\
&\quad\,+\tau ^2 v_0^2 (4 A(0,1,-1)+A(0,1,0))\\
&\quad\,-4 \tau  v_0 (A(1,0,0)+A(1,2,-1))),
\end{split}
\\
\begin{split}
\alpha_{10} &= \frac{1}{2} \tau(4 A(0,1,0) (8 D_\mathrm{T} \tau  (A(0,1,-1)+A(0,1,0))\\
&\quad\,+3 A(2,1,-1))+\tau ^2 v_0^2(8 A(0,1,-1)^2\\
&\quad\,+A(0,1,-1)(42 A(0,1,0)+A(0,1,1))\\
&\quad\,+6 A(0,1,0)^2-A(0,1,1)^2)\\
&\quad\,+\tau v_0(-A(0,1,1) (A(1,2,-2)-2 A(1,2,0))\\
&\quad\,-A(0,1,0) (8 A(1,0,-1)+24 A(1,0,0)\\
&\quad\,+8 A(1,0,1)+3 A(1,2,-2)+36 A(1,2,-1)\\
&\quad\,-2 A(1,2,0))-A(0,1,-1) (8 A(1,0,0)\\
&\quad\,+4 A(1,2,-1)+A(1,2,0)))+A(1,2,0)\\
&\quad\,\quad\,(A(1,2,-2)-A(1,2,0)) +2 A(2,1,0)\\
&\quad\,\quad\,(A(0,1,-1)+3 A(0,1,0)) ),  
\end{split}\raisetag{20ex}\\
\begin{split}
\alpha_{11} &= -4 \tau ^2 A(0,1,0)(\tau  v_0 (16 A(0,1,-1)^2\\
&\quad\, +2 A(0,1,-1)(17 A(0,1,0)+A(0,1,1)) \\
&\quad\,+(A(0,1,0)-A(0,1,1)) (3 A(0,1,0)\\
&\quad\,+2 A(0,1,1)))-4 (A(0,1,-1)+A(0,1,0))\\
&\qquad\,\,(A(1,0,-1)+2 A(1,0,0)+A(1,0,1))\\
&\quad\,-2 A(1,2,-2)(A(0,1,0)+A(0,1,1)) \\
&\quad\,-8 A(1,2,-1)(A(0,1,-1)+2 A(0,1,0)) \\
&\quad\,+A(1,2,0)(A(0,1,0)+2 A(0,1,1)) ), 
\end{split}\raisetag{14ex}\\
\begin{split}
\alpha_{12} &= 16 \tau ^3 A(0,1,0)^2 (16 A(0,1,-1)^2+A(0,1,-1)\\
&\quad\,\quad\,(A(0,1,1)+17 A(0,1,0)) +(A(0,1,0)\\
&\quad\,-2 A(0,1,1))(A(0,1,0)+A(0,1,1))),
\end{split}\raisetag{6.5ex}\\
\begin{split}
\alpha_{13} &= -\frac{1}{2} \tau  v_0(16 D_\mathrm{T} \tau A(0,1,0)+2A(2,1,0)\\
&\quad\,+\tau ^2 v_0^2 (2 A(0,1,-1)+2 A(0,1,0)\\
&\quad\,+3 A(0,1,1))-\tau  v_0 (2 A(1,0,-1)\\
&\quad\, +2 A(1,0,0)+2 A(1,0,1)+3 A(1,2,0))),
\end{split}\\
\begin{split}
\alpha_{14} &= \tau(32 D_\mathrm{T} \tau  A(0,1,0) (A(0,1,-1)\\
&\quad\,+A(0,1,0))+\tau ^2 v_0^2(4 A(0,1,-1)^2 \\
&\quad\,+A(0,1,-1)(24 A(0,1,0)+5 A(0,1,1)) \\
&\quad\,+14 A(0,1,0)^2+32 A(0,1,0) A(0,1,1)) \\
&\quad\,+A(0,1,1)^2-\tau  v_0(A(0,1,1) (A(1,2,-2)\\
&\quad\,+2 A(1,2,0))+A(0,1,-1) (2 A(1,0,-1)\\
&\quad\,+4 A(1,0,0)+2 A(1,0,1)+5 A(1,2,0))\\
&\quad\,+A(0,1,0) (22 A(1,0,-1)+8 A(1,2,-1)\\
&\quad\,+12 A(1,0,0)+22 A(1,0,1)+3 A(1,2,-2)\\
&\quad\,+18 A(1,2,0)))+A(1,2,-2) A(1,2,0)\\
&\quad\,+A(1,2,0)^2+12 A(0,1,0) A(2,1,-1)\\
&\quad\,+2 A(0,1,-1) A(2,1,0)
+6 A(0,1,0) A(2,1,0)),
\end{split}\raisetag{22ex}\\
\begin{split}
\alpha_{15} &= -8 \tau ^2 A(0,1,0)(\tau  v_0 (8 A(0,1,-1)^2\\
&\quad\,+A(0,1,-1)(21 A(0,1,0)+11 A(0,1,1)) \\
&\quad\,+8 A(0,1,0)^2+26 A(0,1,1) A(0,1,0)\\
&\quad\,+2A(0,1,1)^2)-A(0,1,-1) (6 A(1,0,-1)\\
&\quad\,+4 A(1,0,0)+6 A(1,0,1)+4 A(1,2,-1)\\
&\quad\,+5 A(1,2,0))-2 (A(0,1,1) (A(1,2,-2)\\
&\quad\,+A(1,2,0))+A(0,1,0) (5 A(1,0,-1)\\
&\quad\,+2 A(1,0,0)+5 A(1,0,1)+A(1,2,-2)\\
&\quad\,+2 A(1,2,-1)+3 A(1,2,0)))), 
\end{split}\\
\begin{split}
\alpha_{16} &= 32 \tau ^3 A(0,1,0)^2(8 A(0,1,-1)^2 +11 A(0,1,-1)\\
&\quad\,\quad\,(A(0,1,0)+A(0,1,1)) +3 A(0,1,0)^2\\
&\quad\,+2 A(0,1,1)^2+13 A(0,1,0) A(0,1,1)),
\end{split}\raisetag{6.5ex}\\
\begin{split}
\alpha_{17} &= \frac{\tau}{2} (\tau ^2 v_0^2 (24 A(0,1,1) A(0,1,0)+A(0,1,1)^2\\
&\quad\,+7 A(0,1,0)^2+32 A(0,1,-1) A(0,1,0))\\
&\quad\,-2 \tau  v_0(A(0,1,1) A(1,2,0)+A(0,1,0)\\
&\qquad\,\, (8 A(1,0,-1)+8 A(1,0,0)+8 A(1,0,1)\\
&\quad\,+8 A(1,2,-1)+6 A(1,2,0))) +A(1,2,0)^2),  
\end{split}\raisetag{11ex}\\
\begin{split}
\alpha_{18} &= 2 \tau ^2 A(0,1,0)(-\tau  v_0 (32 A(0,1,-1)^2\\
&\quad\,+A(0,1,-1)(103 A(0,1,0)+21 A(0,1,1)) \\
&\quad\,+15 A(0,1,0)^2+71 A(0,1,0) A(0,1,1))\\
&\quad\,+6 A(0,1,1)^2+4A(0,1,1) A(1,2,-2)\\
&\quad\,+8 A(0,1,-1) (A(1,0,-1)+2 A(1,0,0)\\
&\quad\,+A(1,0,1)+A(1,2,-1))+4A(0,1,0) \\
&\quad\,\quad\,(6 A(1,0,-1)+4 A(1,0,0)+6 A(1,0,1)\\
&\quad\,+A(1,2,-2)+6 A(1,2,-1))+A(1,2,0)\\
&\quad\,\quad\,(6 A(0,1,1)+13 A(0,1,0)+9 A(0,1,-1)) ), 
\end{split}\raisetag{16.5ex}\\
\begin{split}
\alpha_{19} &= 64 \tau ^3 A(0,1,0)^2(8 A(0,1,-1)^2+A(0,1,0)^2\\
&\quad\, +A(0,1,1)^2+A(0,1,-1)(9 A(0,1,0)\\
&\quad\, +5 A(0,1,1)) +6 A(0,1,0) A(0,1,1)).
\end{split}
\end{align}
Those in Eq.\ \eqref{eqn:4thorderP} are given by
\begin{align}
\beta_{1} &= -\tau v_0,\\
\beta_{2} &= 8\tau A(0,1,0),\\
\beta_{3} &= -\frac{1}{16} \tau ^2 v_0 (16 D_\mathrm{T}+\tau  v_0^2),\\
\begin{split}
\beta_{4} &= \frac{1}{4} \tau  (32 D_\mathrm{T} \tau  A(0,1,0)+6 A(2,1,0)+\tau  v_0 (\tau  v_0 \\
&\quad\,\quad\,\,\,(4 A(0,1,0)+A(0,1,1))-4A(1,0,-1)\\
&\quad\, -4A(1,0,1) -4A(1,2,-1)-A(1,2,0))),
\end{split}\raisetag{6.5ex}\\
\begin{split}
\beta_{5} &= \tau ^2 A(0,1,0) (-\tau  v_0 (5 A(0,1,0)\\
&\quad\,+3 A(0,1,1))+8A(1,0,-1)+8A(1,0,1)\\
&\quad\,+8A(1,2,-1)+A(1,2,0)),
\end{split}\\
\beta_{6} &= 8 \tau ^3 A(0,1,0)^2 (A(0,1,0)+A(0,1,1)),\\
\begin{split}
\beta_{7} &= \frac{1}{4} \tau  (2 A(2,1,0)+\tau  (32 D_\mathrm{T} A(0,1,0)\\
&\quad\,+\tau  v_0^2 (8 A(0,1,-1)+2 A(0,1,0)\\
&\quad\,+A(0,1,1))-v_0 (8 A(1,0,0)\\
&\quad\,+4 A(1,2,-1)+A(1,2,0)))),
\end{split}\\
\begin{split}
\beta_{8} &= -4 \tau ^2 A(0,1,0) (\tau  v_0 (8 A(0,1,-1)\\
&\quad\,+A(0,1,0)+A(0,1,1))-2A(1,0,-1)\\
&\quad\,-4 A(1,0,0)-2A(1,0,1)-4 A(1,2,-1)),
\end{split}\\
\begin{split}
\beta_{9} &= 8 \tau^3 A(0,1,0)^2 (16 A(0,1,-1)+A(0,1,0)\\
&\quad\,+A(0,1,1)),
\end{split}\\
\begin{split}
\beta_{10} &= \frac{1}{2} \tau  (\tau  (32 D_\mathrm{T} A(0,1,0)+v_0 (\tau  v_0 (4A(0,1,-1)\\
&\quad\,+4A(0,1,0)+5 A(0,1,1))-2A(1,0,-1)\\
&\quad\,-4 A(1,0,0) -2A(1,0,1)
-5 A(1,2,0)))\\
&\quad\, +2 A(2,1,0)),
\end{split}\raisetag{8ex}\\
\begin{split}
\beta_{11} &= -4 \tau ^2 A(0,1,0) (\tau  v_0 (8 A(0,1,-1)+5 A(0,1,0)\\
&\quad\, +11 A(0,1,1))-6 A(1,0,-1)-4 A(1,0,0)\\
&\quad\,-6 A(1,0,1)-4 A(1,2,-1)-5 A(1,2,0)),
\end{split}\raisetag{6.3ex}\\
\begin{split}
\beta_{12} &= 16 \tau ^3 A(0,1,0)^2 (8 A(0,1,-1)+3 A(0,1,0)\\
&\quad\,+11 A(0,1,1)),
\end{split}\\
\begin{split}
\beta_{13} &= \tau^2 A(0,1,0)(-\tau v_0(32 A(0,1,-1)\\ 
&\quad\, +7A(0,1,0) +21 A(0,1,1))+8 (A(1,0,1)\\
&\quad\, +2A(1,0,0)+A(1,0,-1)+A(1,2,-1)) \\
&\quad\, +9 A(1,2,0)),
\end{split}\\
\begin{split}
\beta_{14} &= 32 \tau^3 A(0,1,0)^2 (8 A(0,1,-1)+A(0,1,0)\\
&\quad\,+5 A(0,1,1)).
\end{split}
\end{align}
The coefficients in Eq.\ \eqref{eqn:4thorderQ} are given by
\begin{align}
\gamma_1 &= \frac{1}{8} \tau ^2 v_0^2,\\
\gamma_2 &= -\frac{1}{2} \tau  (\tau  v_0 (3 A(0,1,0)+A(0,1,1))-A(1,2,0)),\\
\gamma_3 &= 4 \tau ^2 A(0,1,0) (A(0,1,0)+A(0,1,1)).
\end{align}

\nocite{apsrev41Control}
\bibliographystyle{apsrev4-1}
\bibliography{control,refs}

\end{document}